# Multiscale Graph Neural Network for Turbulent Flow-Thermal Prediction Around a Complex-Shaped Pin-Fin

Riddhiman Raut[1, #], Evan M. Mihalko[1, #], Amrita Basak[1, *]

[1]Department of Mechanical Engineering, Pennsylvania State University, University Park, PA 16802, United States

\# Equal contribution

∗Corresponding author aub1526@psu.edu (A. Basak)

## Abstract

*This study presents the development of a domain-responsive edge-aware multiscale Graph Neural Network for predicting steady, turbulent flow and thermal behavior in a two-dimensional channel containing arbitrarily shaped complex pin-fin geometries. The training dataset was constructed through an automated framework that integrated geometry generation, meshing, and flow-field solutions in ANSYS Fluent. The pin-fin geometry was parameterized using piecewise cubic splines, producing 1,000 diverse configurations through Latin Hypercube Sampling. Each simulation was converted into a graph structure, where nodes carried a feature vector containing spatial coordinates, a normalized streamwise position, one-hot boundary indicators, and a signed distance to the nearest boundary (e.g., wall). This graph structure served as input to the newly developed Graph Neural Network, which was trained to predict temperature, velocity magnitude, and pressure at each node using data from ANSYS. The network predicted fields with outstanding accuracy – capturing boundary layers, recirculation, and the stagnation region upstream of the pin-fins – while reducing wall time by 2–3 orders of magnitude. In conclusion, the novel graph neural network offered a fast and reliable surrogate for simulations in complex flow configurations.*

## 1. Introduction

Accurately predicting fluid flow around complex geometries is a central task in computational fluid dynamics (CFD), with critical applications in the aerospace, automotive, and semiconductor industries[1–3]. In these domains, flow behavior often dictates performance, safety, and efficiency, making high-fidelity simulation tools essential to both the design and optimization process[4]. Traditional CFD solvers, while powerful, are computationally intensive, often requiring the use of supercomputers, especially when applied to large-scale or high-resolution simulations[5]. These tools operate by numerically solving the governing partial differential equations over a discretized domain. This process involves generating a mesh that captures the geometry, discretizing the governing equations, and solving the resulting system of equations iteratively until convergence is achieved. The computational cost of traditional CFD increases significantly with finer meshes, more complex geometries, multi-physics coupling (e.g., thermal-hydraulics or reactive flows), or when precise tolerances on convergence are required[6]. These challenges become particularly prohibitive for design exploration, optimization, uncertainty quantification, or real-time control scenarios, where simulations must be repeated across a wide range of geometries or operating conditions[7–9]. In such contexts, the computational burden of traditional solvers limits the speed and scalability of the engineering design cycle.

In response to the high computational costs and iterative nature of traditional CFD, the fluid mechanics community has increasingly explored scientific machine learning (SciML) techniques as an effective means to accelerate fluid simulations and enable real-time design and optimization. Initially, SciML applications in CFD primarily involved classical surrogate modeling methods, such as Gaussian processes, polynomial chaos expansions, and various regression models, demonstrating effective predictive capabilities for relatively low-dimensional fluid flow problems, including simplified laminar flows around cylinders or basic aerofoils[10,11]. However, these early methods were constrained by their inability to scale effectively to high-dimensional and turbulent flow scenarios.

The advent of deep learning significantly expanded the capabilities of SciML in fluid dynamics. Architectures such as convolutional neural networks (CNNs), recurrent neural networks (RNNs), and autoencoders have been successfully applied to predict fluid and thermal behaviors, notably in structured-grid contexts. CNN-based surrogate models, in particular, have shown notable success in tasks ranging from turbulence modeling and heat transfer prediction to physics-informed surrogate modeling, providing accelerated yet reasonably accurate alternatives to classical CFD methods. For instance, Morimoto et al.[12] successfully analyzed Particle Image Velocimetry (PIV) data using CNNs, while Kochkov et al.[13] applied end-to-end CNN models to improve approximations within CFD domains for modeling of two-dimensional flows. Similarly, convolutional autoencoders trained with physics-informed regularizations have shown success in data compression, with no adversely affecting reconstruction quality[14]. Despite their effectiveness, these traditional deep learning methods inherently rely on structured grid data, severely limiting their applicability to industrial CFD problems typically characterized by complex geometries and irregular, unstructured meshes.

As a result, graph neural networks (GNNs) have emerged as a promising class of models for accelerating CFD simulations[15–17]. While conventional deep learning models like CNNs are designed for data structured on regular grids, GNNs are inherently suited for data defined over irregular domains, such as unstructured meshes[18]. This characteristic makes GNNs particularly attractive for CFD, where the underlying discretization schemes of commercial solvers are often complex and nonuniform in nature. By operating directly on graph-based representations of the computational domain, GNNs can naturally incorporate geometric and topological information, enabling the model to learn localized physical interactions such as boundary layer development, and propagate information across the domain. Moreover, GNNs offer the flexibility to handle varying mesh resolutions and topologies without requiring explicit interpolation or remeshing steps. This capability is critical for surrogate modeling in engineering applications, where the need to simulate flows over a wide range of geometries and boundary conditions demands models that can generalize beyond a fixed spatial discretization. Landmark studies by Battaglia et al.[19] and Pfaff et al.[15] have demonstrated the foundational capability of GNNs to capture fluid dynamics directly from simulation-generated datasets, effectively predicting transient fluid flows with impressive accuracy and robustness. More recent advancements, such as those by Sanchez-Gonzalez et al.[20], have further established the efficacy of GNNs in modeling turbulent fluid phenomena, explicitly demonstrating their ability to generalize across varying flow conditions and geometries, and predict complex transient behaviors. Chen et al.[21] also successfully applied GNNs to model laminar flow around a variety of two-dimensional objects.

However, several critical limitations persist, especially in capturing intricate flow phenomena and generalizing predictions across a variety of geometries. Traditional GNN architectures often struggle to accurately resolve near-wall behaviors, including boundary layers, recirculation zones, and flow separation regions[22]. For example, Barwey et al.[23] introduced interpretable error-tagging modules to baseline GNNs, aiming to identify and reduce prediction errors specifically in complex flow regions. Nevertheless, this approach requires a fully trained baseline GNN and substantial additional training of the interpretability module, effectively doubling the training time. Alternatively, Travnikov et al.[24] addressed the issue of GNNs' limited receptive fields by incorporating global pooling operations and a fully connected branch parallel to the primary GNN layers. Although this method improved overall performance, significant errors persisted, particularly in wake regions and near leading-edge boundaries of test geometries such as airfoils. Moreover, many existing GNN-based CFD models rely heavily on specific datasets, severely restricting their ability to generalize to unseen geometrical configurations[25–27]. As a consequence, models that accurately predict flows around common shapes like NACA airfoils often falter when applied to novel or complex geometries not represented in their training sets[24]. In short, current GNN surrogates either (i) incur large errors in boundary dictated regions, (ii) fail to extrapolate beyond narrowly curated training meshes, or (iii) demand two stage training pipelines that negate their touted speedups. These persistent shortcomings indicate that purely architectural modifications, without careful integration of physical and geometric information, may be inadequate – or at least inefficient – in achieving robust, generalizable, and highly accurate predictions necessary for real-world CFD applications.

To bridge this gap, the present study introduced domain-responsive edge-aware multiscale Graph Neural Network (DREAM-GNN), which explicitly integrated boundary-aware node and edge features, employed multiscale hierarchical message-passing, and provided an adaptive framework designed specifically to resolve boundary-driven flow phenomena and generalize effectively across diverse geometries. Two-dimensional pin-fins were selected as the domain of choice – pin-fin arrays are ubiquitous in gas turbine blade cooling passages and thermal management hardware, hence, accelerating design space exploration translates directly into shorter design cycle times and rapid convergence of optimization frameworks. Additionally, these geometries involve complex flow features such as boundary layer separation, wake interactions, and stagnation zones, making them an ideal benchmark for evaluating the ability of surrogate models to capture multi-scale physics. To begin with, complex pin-fin geometries were parameterized using four connecting piecewise cubic splines, providing control over the shape and orientation. The training data was generated through an automated Design-of-Experiments (DoE) framework utilizing ANSYS Fluent, incorporating best practices such as mesh refinement and validated viscous modeling. The automated framework extracted key information such as mesh connectivity and associated field quantities to be used for training in predicting the temperature, velocity magnitude, and pressure fields. The predictive performance of DREAM-GNN was rigorously evaluated and compared against popular state-of the-art models such as GCN[28,29], GraphSAGE[30]. Results demonstrated that the developed DREAM-GNN achieved substantial acceleration over traditional CFD solvers but also surpassed existing SciML approaches in accuracy, robustness, and generalization to new designs.

The remainder of the manuscript is structured as follows: Section 2 discusses simulation methodology, data generation, GNN model development, and data preprocessing. The observations and ensuing discussion are delineated in Section 3, while Section 4 summarizes the conclusions observed.

## 2. Methods

This section begins with the methodology for generating arbitrarily shaped pin-fins, followed by a detailed description of the computational domain. Subsequently, the procedures used to produce ground truth simulation data are outlined. Finally, the development and design of the GNN models are presented.

### 2.1 Generation of complex pin-fin shapes

The pin-fin was constructed using four connecting spline segments, each defined within a partitioned angular domain[31]. To achieve this, the angular positions of the spline endpoints were first determined by discretizing a full revolution ($2\pi$) into four equal partitions, resulting in a vector of angles, $\theta_n$, where $\theta_n = \frac{2\pi}{4}n$ for n = 0, 1, 2, 3. To introduce rotational flexibility, a global rotation angle, $\theta_0$, was added to each $\theta_n$, producing the final set of angular positions that defined the shape's orientation. The radial coordinates at each of these angular positions were specified by a vector $r_n = [r_1, r_2, r_3, r_4]$, where each $r_n$ represented the radial distance from the center to the shape's boundary at the corresponding $\theta_n$. These four control points – each defined by $(r_n, \theta_n + \theta_0)$ in polar coordinates – served as the basis for constructing the shape using spline interpolation, ensuring a smooth transition between adjacent segments. This formulation allowed for a compact, parametric representation of the shape while maintaining control over both its geometric proportions and its orientation within the domain. The individual parametric functions $r_n \cos(\theta_n)$ and $r_n \sin(\theta_n)$, that defined the boundary in Cartesian coordinates, are shown in Fig. 1(a). The resulting closed shape, obtained by interpolating between the control points with splines, is depicted in Fig. 1(b). Figs. 1(c) and 1(d) illustrate a second shape generated using the same framework, but with a different set of $r_n$ values and a modified rotation angle, $\theta_0$, demonstrating the flexibility of the parametric design. The individual parameters and their respective bounds used in this formulation are listed in Table 1. It is important to note that $r_1$ remained fixed at a constant value for all combinations to avoid mathematical errors in the spline generation.

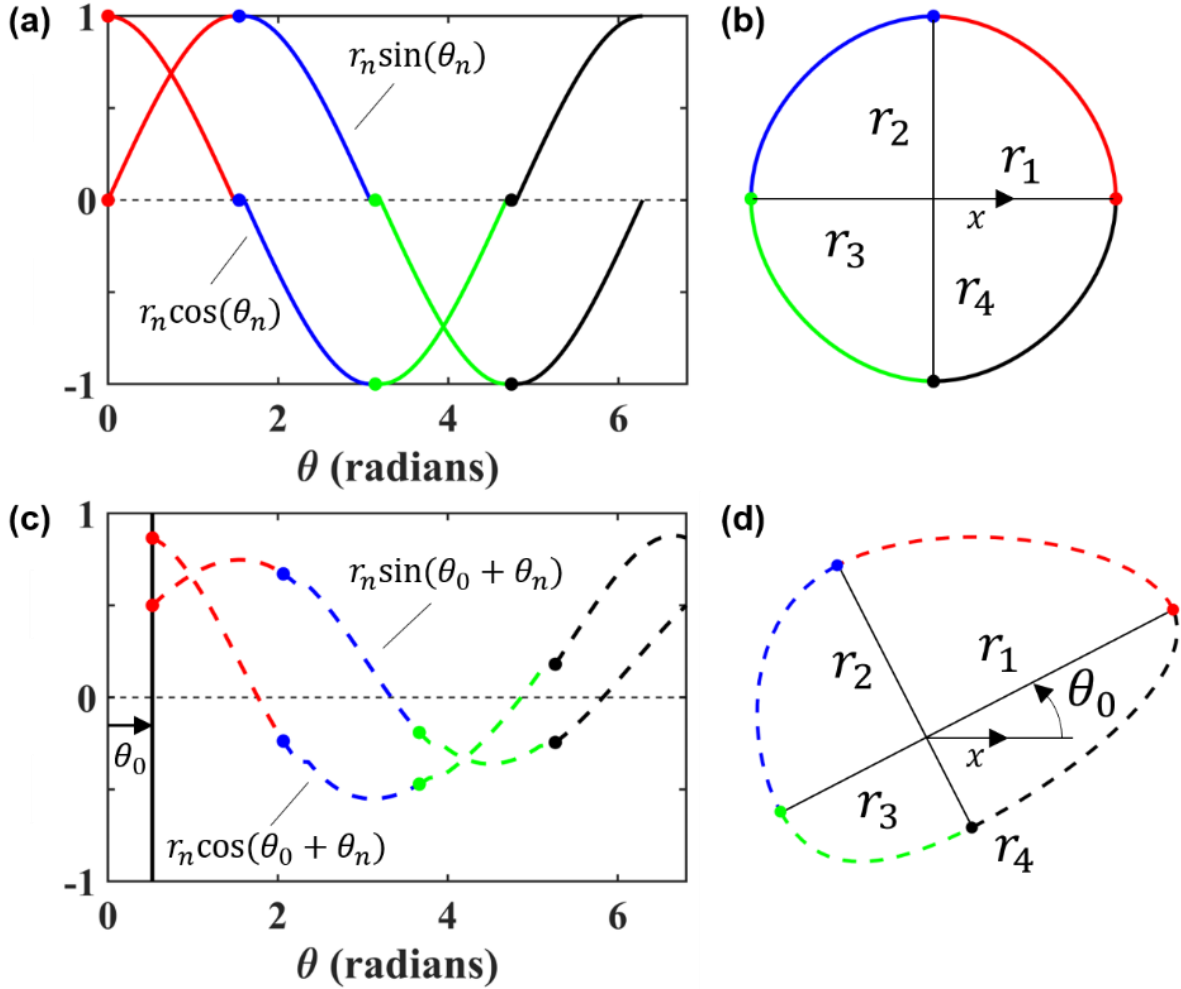

**Fig. 1.** Parametric construction of the pin-fin shape. (a) Polar coordinates defining the control points based on the radial parameters $r_n$ and angular position $\theta_n$ resulting in (b) closed pin-fin shape. (c) Example of an alternative shape generated using a different set of radial parameters and a change in orientation angle represented by, $\theta_0$, resulting in a different closed pin-fin shape in (d).

**Table 1:** Geometric input parameters and bounds for pin-fin shape generation

| Parameter | Description | Bounds |
|---|---|---|
| $r_1$ (mm) | Radius length at $r_1$ | 1 |
| $r_2$ to $r_4$ (mm) | Radius length at $r_2$ to $r_4$ | [0.1, 1] |
| $\theta_0$ (radians) | Orientation angle | [0, π] |

## 2.2 Description of the two-dimensional domain

This study investigated fluid flow and heat transfer behavior within a two-dimensional channel geometry measuring 15 mm in length (L) and 6 mm in width (W). A single pin-fin structure was introduced into the flow domain to explore its influence on local flow dynamics and thermal performance. The nominal position of the pin-fin was fixed at 5 mm downstream from the channel inlet, aligned along the channel's centerline to ensure symmetric boundary conditions and flow development under idealized circumstances.

However, the pin-fins analyzed in this study were not limited to a single geometry; instead, they were generated using a shape-generation framework designed to explore a wide morphological space. As a result of this generative process, many of the resulting pin-fin geometries were asymmetric in shape, which often caused slight shifts in their effective center of mass or spatial footprint within the channel. This led to deviations from the originally intended central location, introducing geometric variability that more accurately reflects practical design irregularities or intentionally perturbed designs.

Fig. 2 provides a schematic of the overall domain, illustrating the nominal flow path and pin-fin placement. The inset table within the figure shows several examples of the generated pin-fin geometries, along with the corresponding parametric inputs (in mm) used to define their shapes. This dataset of diverse pin-fin configurations enables a broad exploration of how geometric features – such as asymmetry, elongation, and surface curvature – influence flow structures and heat transfer performance within a constrained channel.

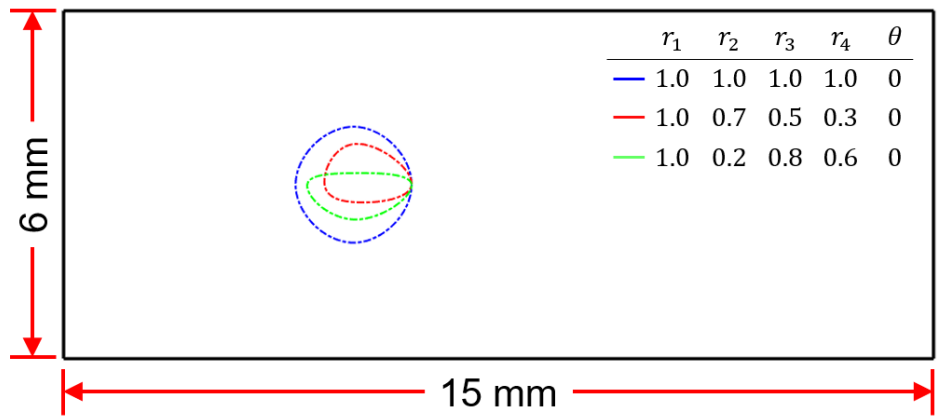

**Fig. 2.** Schematic representation of the two-dimensional pin-fin domain with representative pin-fins. The pin-fin shapes were generated using the parameters listed on the figure. The variables $r_1$ to $r_4$ are in mm.

## 2.3 Development of a computational model

### 2.3.1 Governing equations and boundary conditions

The flow and thermal fields within the two-dimensional domain were governed by the steady-state Reynolds-Averaged Navier-Stokes (RANS) equations, which model the conservation of mass, momentum, and energy for turbulent flows. The continuity equation ensured mass conservation throughout the domain. The momentum equations accounted for the balance of forces, including pressure gradients, viscous stresses, and Reynolds stresses arising from turbulence. The energy equation described the transport of thermal energy, incorporating conduction and convective heat transfer, with effective thermal properties adjusted to capture turbulence effects. To simplify the problem and improve computational efficiency while retaining physical accuracy, several assumptions were made: the flow was steady and two-dimensional; air was treated as the working fluid with constant thermophysical properties evaluated at 300 K; no-slip boundary conditions were applied at solid surfaces including the pin-fins and walls; and gravitational and radiative heat transfer effects were neglected.

For the two-dimensional domain, velocity-inlet and pressure-outlet boundary conditions were imposed. The flow regime corresponded to a Reynolds number of ~10,000, with an inlet air temperature of 300 K, calculated using the two-dimensional hydraulic diameter, $D_h$. The outlet was maintained at a gauge pressure of zero, representing air exhausting into the ambient environment. The side walls and pin-fin were subjected to a no-slip condition and held at a constant isothermal temperature of 350 K. The inlet region was extended upstream of the pin-fin to ensure uniform flow upon initial contact. Similarly, the outlet region was extended downstream to capture the wake region and allow for flow stabilization before exiting the domain, minimizing numerical errors. A schematic representation of the boundary conditions is shown in Fig. 3 with detailed specifications listed in Table 2.

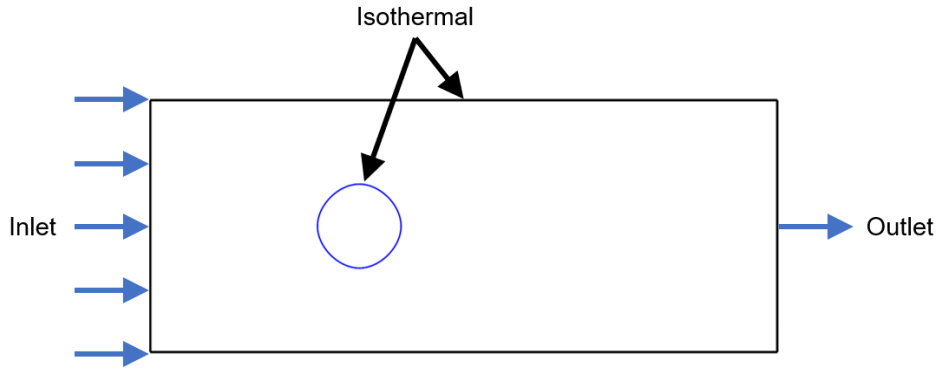

**Fig. 3.** Schematic of the simulation domain with boundary conditions

**Table 2:** Detailed specifications of domain boundary conditions in ANSYS Fluent

| Boundary | Boundary Conditions | Values |
|---|---|---|
| Inlet | Constant velocity | 15.51 m/s |
| | Turbulent intensity | 5% |
| | Hydraulic diameter | 0.012 m |
| | Thermal | 300 K |
| Outlet | Pressure outlet | Zero gauge pressure |
| | Turbulent intensity | 5% |
| | Hydraulic diameter | 0.012 m |
| | Thermal | 300 K |
| Pin-fin | No-slip – stationary wall | - |
| | Isothermal | 350 K |
| Side walls | No-slip – stationary wall | - |
| | Isothermal | 350 K |

### 2.3.2 Meshing and turbulence model

In this study, ANSYS Fluent was used to generate a triangular mesh with linear-order elements for the two-dimensional domain. The overall mesh element size was set to 0.05 mm, with refinement applied at the pin-fin and side walls, where the element size was reduced to 0.005 mm. This meshing strategy ensured accurate resolution of the boundary layer and thermal gradients around the pin-fin, which were critical for reliable flow and heat transfer predictions. The refined mesh near the pin-fin captured the detailed flow behavior, while the coarser mesh in the far-field regions balanced computational efficiency. To ensure the robustness of the numerical results, a grid independence study was conducted by systematically decreasing the overall element size and increasing the level of refinement at the walls, which correspondingly increased the node count.

The resulting trends in heat transfer ($Q$) and pressure drop ($\Delta P$) are shown on Fig. 4(a) and 4(b), respectively as functions of the total number of nodes. Beyond the selected mesh, identified by a red dot in both plots, changes in these quantities were less than 2%, indicating convergence of the solution and validating the mesh choice used throughout this study. A representative sample mesh is shown on Fig. 4(c), illustrating the entire computational domain and the local refinements around an arbitrarily selected pin-fin shape. The final mesh consisted of approximately 42,000 nodes and 83,000 elements; however, with each new design, the number of nodes and elements varied based on the pin-fin shape. The SST k-ω turbulence model was employed due to its proven accuracy in resolving near-wall effects and handling adverse pressure gradients[32]. With the selected mesh resolution, the average $Y^+$ at the pin-fin and walls was approximately 0.12, ensuring that the first cell remained well within the viscous sublayer for accurate turbulence modeling, with all values not exceeding 1[33,34].

This mesh configuration resulted in an average simulation completion time of approximately 8 minutes using a system equipped with an Intel Xeon 5217 CPU, RTX 4000 GPU, and 192 GB of memory. These simulations provided the ground truth data essential for training the GNN model, capturing critical flow and thermal characteristics necessary for robust model performance.

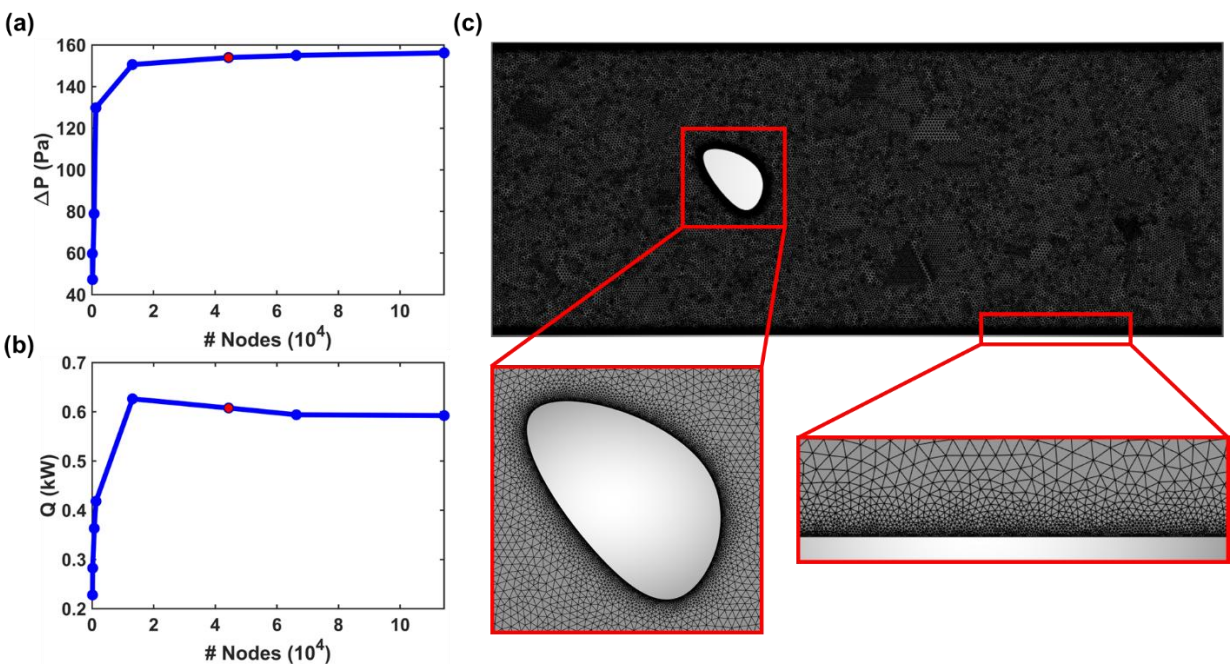

**Fig. 4.** Mesh convergence study with (a) pressure drop and (b) heat transfer. (c) Representative two-dimensional mesh using triangulation and refinement at the pin-fin and side walls

## 2.3 Dataset generation for training

The training dataset was generated using an enhanced version of an automated workflow established in prior literature[31,35], which integrates multiple software tools. Each pin-fin shape was parameterized by the geometric input variables and bounds listed in Table 1. While $r_1$ was an input parameter, it was fixed at 1 mm across all designs to prevent numerical errors in the shape generation and was therefore not included as a varying parameter. The design space was explored using a Latin Hypercube Sampling (LHS) approach to generate 1,000 unique pin-fin geometries, ensuring a diverse and well-distributed dataset. An automated pipeline linked key processes, enabling seamless geometry generation, meshing, and CFD solution evaluation using ANSYS Fluent. This iterative process produced simulation results that accurately captured the flow and thermal characteristics of the system. From the mesh, individual node coordinates and connectivity were extracted, along with solution data, including temperature, pressure, and velocity values at each node location. A schematic of the automated pipeline used to generate the training data is shown in Fig. 5.

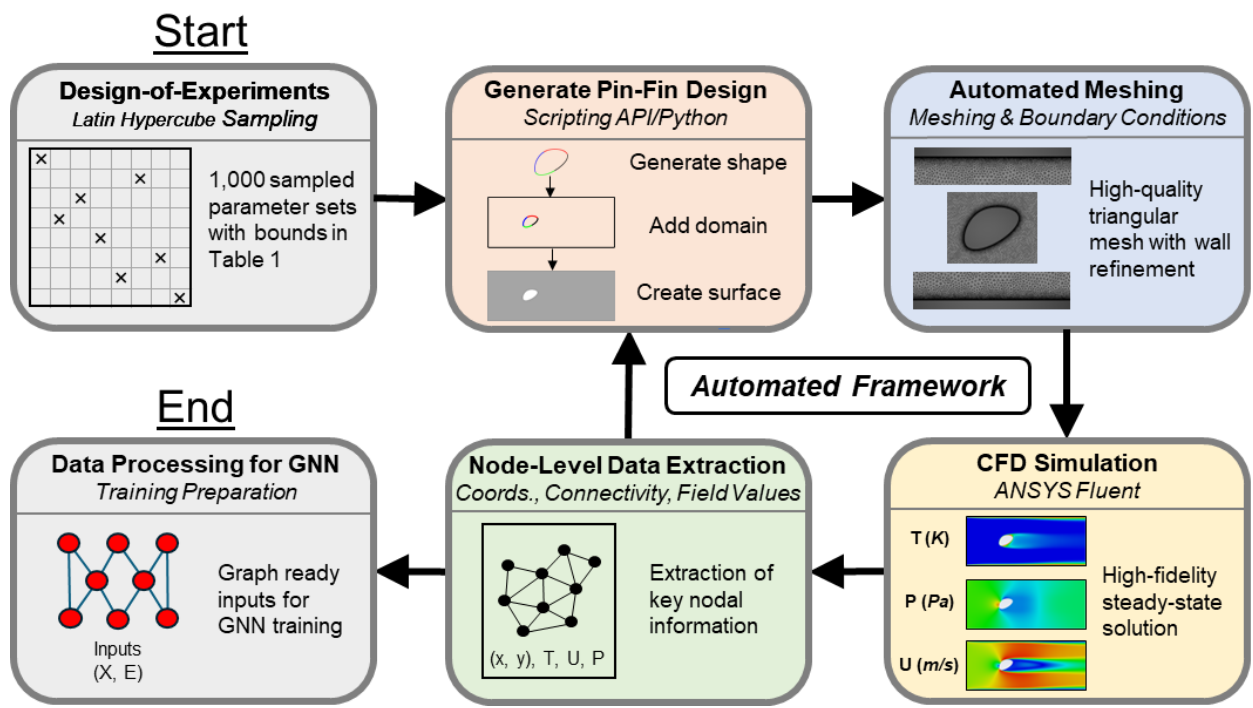

**Fig. 5.** Schematic of the dataset generation pipeline for model training. The process begins with LHS generating 1,000 parameter sets within the bounds specified in Table 1. For each set, a corresponding pin-fin geometry and domain are created. The domain is then meshed and simulated in ANSYS Fluent. Upon completion, nodal information and solution field values are extracted and then processed for use in model training.

### 2.4 Development of the Graph Neural Network (GNN) model

#### 2.4.1 Graph construction and featurization

Each CFD simulation was converted into a graph-based representation suitable for processing by a multiscale GNN. Each finite-volume cell center in the CFD mesh became a node $v_i \in \mathcal{V}$, located at physical coordinates $x_i = (x_i, y_i)$. Edges connected nodes whose corresponding cells shared a common face, preserving the original mesh connectivity. To ensure symmetric message passing, edges were recorded bidirectionally. A typical graph contains approximately $4.23 \times 10^4$ nodes and $1.25 \times 10^5$ edges, with moderate variation across the dataset of 1,000 graphs.

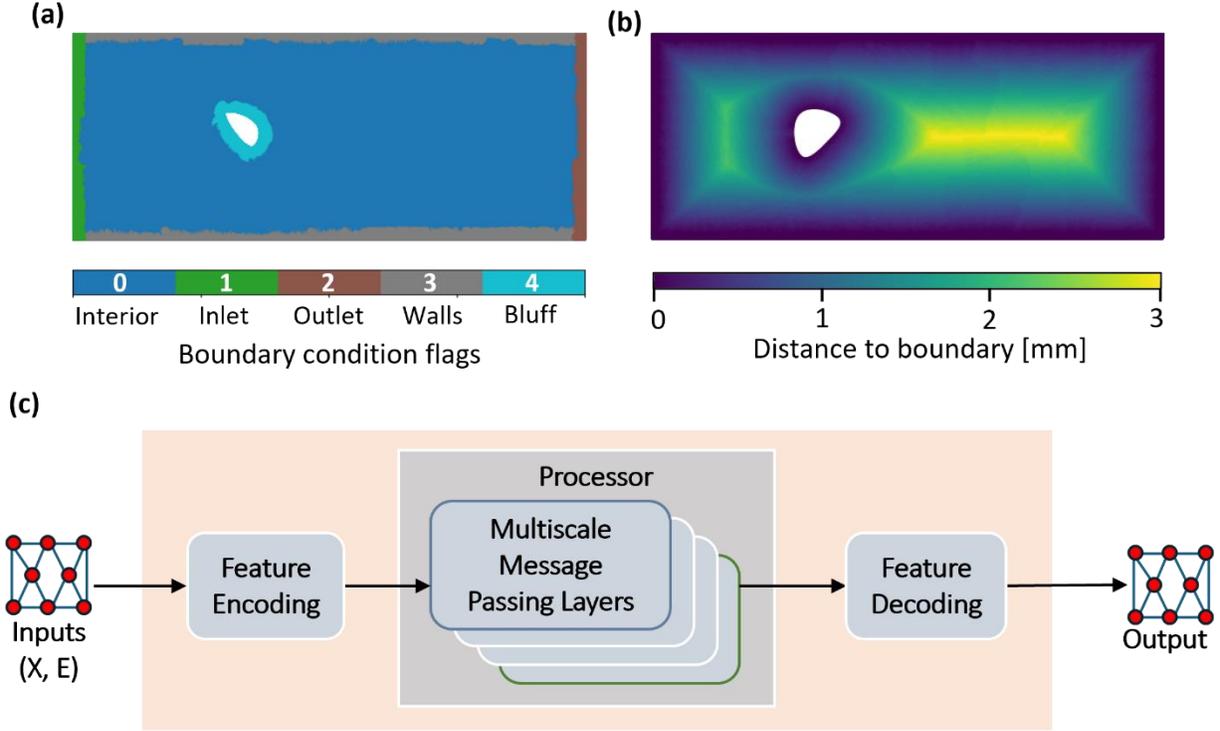

**Fig. 6.** Feature engineering and model architecture. The one-hot boundary encoding scheme is shown in (a), distinguishing interior fluid (0), inlet (1), outlet (2), side walls (3), and pin-fin walls (4). Panel (b) demonstrates the signed-distance field from the nearest solid boundary, giving the network explicit geometric context. Panel (c) shows the model with an encoder-processor-decoder architecture, with the processor consisting of 4 multiscale message passing layers.

Each node $v_i$ carried a nine-dimensional feature vector $h_i$, which included spatial coordinates $(x_i, y_i)$, a normalized streamwise coordinate $\tilde{x}_i$, a one-hot boundary encoding, and a signed distance to the nearest boundary. The streamwise coordinate was computed per graph as

$$\tilde{x}_i = \frac{x_i - x_{min}}{x_{max} - x_{min}} \tag{1}$$

providing an explicit representation of flow progression from inlet ($\tilde{x}_i = 0$) to outlet ($\tilde{x}_i = 1$). Boundary conditions were encoded using a five-dimensional one-hot vector indicating whether a node belonged to the interior fluid (0), inlet (1), outlet (2), wall (3), or bluff-body surface (4), as illustrated in Fig. 6(a). The signed distance $d_i$ was computed as the Euclidean distance to the closest boundary point (Fig. 6(b)). Also, each directed edge $e_{ij}$ carried a four-dimensional feature vector $g_{ij}$, defined as

$$g_{ij} = [\Delta x_{ij}, \Delta y_{ij}, ||\Delta x_{ij}||_2, sign(\Delta x_{ij})]^\top, \quad \Delta x_{ij} = x_j - x_i \tag{2}$$

The first two components, $\Delta x_{ij}$ and $\Delta y_{ij}$, represent the relative spatial displacement between nodes and encode geometric orientation. The third component, $||\Delta x_{ij}||_2$, captures the Euclidean distance between nodes, which provides a local length scale for multiscale filtering. The final component, $sign(\Delta x_{ij})$, indicates whether the edge points downstream or upstream with respect to the flow direction, introducing explicit flow alignment into the message-passing process. To ensure stable learning across the dataset, the geometric features $(x_i, y_i, d_i)$ were normalized using global z-score normalization:

$$x_i^\star = \frac{x_i - \bar{x}}{\sigma_x}, \quad y_i^\star = \frac{y_i - \bar{y}}{\sigma_y}, \quad d_i^\star = \frac{d_i - \bar{d}}{\sigma_d} \tag{3}$$

Here, $\bar{x}$, $\bar{y}$, and $\bar{d}$, are the global means of $x_i$, $y_i$, and $d_i$, respectively, while $\sigma_x$, $\sigma_y$, and $\sigma_d$ are their respective standard deviations. The normalized coordinates $(x_i^\star, y_i^\star, d_i^\star)$ replaced the original values in the node features, while the per-graph streamwise coordinate $\tilde{x}_i$ and one-hot encodings remained unchanged.

The target variables for each node were the steady-state values of temperature $T_i$, pressure $P_i$, and velocity magnitude $U_i$, obtained from the CFD solution. These target values were normalized using global z-score statistics as well, with $\bar{T}$ and $\sigma_T$, $\bar{P}$ and $\sigma_P$, and $\bar{U}$ and $\sigma_U$ representing the global means and standard deviations for $T$, $P$, and $U$, respectively:

$$T_i^\star = \frac{T_i - \bar{T}}{\sigma_T}, \quad P_i^\star = \frac{P_i - \bar{P}}{\sigma_P}, \quad U_i^\star = \frac{U_i - \bar{U}}{\sigma_U} \tag{4}$$

The GNN learned to predict these normalized target values directly from the input graph features. Formally, the network approximated a mapping:

$$f_\theta : (h_i, g_{ij}) \mapsto (T_i^\star, P_i^\star, U_i^\star) \tag{5}$$

Here, $\theta$ denotes the trainable parameters. The model was trained by minimizing the mean squared error (MSE) loss:

$$\mathcal{L}(\theta) = \frac{1}{|\mathcal{V}|} \sum_{i \in \mathcal{V}} \left[ (T_i^\star - \widehat{T_i^\star})^2 + (P_i^\star - \widehat{P_i^\star})^2 + (U_i^\star - \widehat{U_i^\star})^2 \right] \tag{6}$$

Here, $\widehat{T_i^\star}$, $\widehat{P_i^\star}$, and $\widehat{U_i^\star}$ denote the network's predicted values at each node. The dataset was randomly divided into training (70%), validation (15%), and test (15%) sets, ensuring that all subsets were representative of the full range of pin-fin geometries and mesh resolutions. Normalization parameters were stored alongside each graph and applied consistently during training and inference.

### 2.4.2 Model architecture

The model followed a three-stage architecture comprising feature encoding, a processor block made of several multiscale message-passing (MMP) layers, and a final decoding module (Fig. 6(c)). The input node features $h_i$ and edge features $g_{ij}$ were first independently projected into a shared 128-dimensional latent space using two separate multi-layer perceptrons (MLPs), denoted by $\phi_{\text{node}}$ and $\phi_{\text{edge}}$, respectively:

$$h_i^{(0)} = \phi_{\text{node}}(h_i), \quad e_{ij}^{(0)} = \phi_{\text{edge}}(g_{ij}) \tag{7}$$

The encoded graph was then passed through a stack of four Multiscale Message Passing (MMP) layers, each operating over a multiresolution graph hierarchy. Within each MMP layer, message passing was performed at several coarsened levels using seven sequential Message Passing Layer blocks. Each of these blocks updated both edge and node features via MLPs:

$$e_{ij}^{(l)} \leftarrow e_{ij}^{(l-1)} + \phi_{\text{edge}}^{(l)}\left(\left[h_i^{(l-1)}, h_j^{(l-1)}, e_{ij}^{(l-1)}\right]\right) \tag{8}$$

$$h_i^{(l)} \leftarrow h_i^{(l-1)} + \phi_{\text{node}}^{(l)}\left(\left[h_i^{(l-1)}, \sum_{j \in \mathcal{N}(i)} e_{ij}^{(l)}\right]\right) \tag{9}$$

Here, the edge update depends on both the incoming node embeddings and existing edge features, while the node update aggregates neighborhood edge messages and combines them with the current node state. Both update steps applied ReLU activation and layer normalization. Each MMP layer performed hierarchical pooling and unpooling across multiple graph resolutions. The graph was coarsened using voxel-based spatial binning with resolution controlled by a characteristic edge length. After downward passes through increasingly coarse graphs, node features were interpolated back to finer graphs using k-nearest neighbor interpolation (with $k = 4$), followed by upward message-passing to refine the representations. The final output was produced by another MLP decoder $\phi_{\text{decode}}$, which mapped the latent node embeddings $h_i^{\text{fnl}}$ to the predicted physical fields:

$$(\widehat{T_i^\star}, \widehat{P_i^\star}, \widehat{U_i^\star}) = \phi_{\text{decode}}(h_i^{\text{fnl}}) \tag{10}$$

Here, $\widehat{T_i^\star}$, $\widehat{P_i^\star}$, and $\widehat{U_i^\star}$ are the predicted temperature, pressure, and velocity magnitude, respectively, at node $i$. This formulation enabled the model to efficiently aggregate both local and global flow features across nonuniform CFD domains. All updates – including edge and node updates – were performed entirely using learnable MLPs rather than fixed convolutional filters, making the architecture highly expressive and adaptable to mesh-based variations across samples.

All training was performed on a single NVIDIA Tesla V100 GPU hosted on the Roar Collab high-performance computing cluster at Institute for Computational and Data Sciences, the Pennsylvania State University, University Park, Pennsylvania, USA. The network was optimized with the Adam optimizer (initial learning rate $2 \times 10^{-4}$, with a decay of 0.5 every 25 epochs). Training for 100 epochs on the 700graph dataset required a wall time of around 6 hours. Inference was performed on the dual-processor Intel® Xeon® Gold 6230R CPU, operating at a base frequency of 2.1 GHz. The inference time was clocked at less than 1 second for a single graph, ~500 times speedup compared to that of the Fluent RANS solver.

## 3. Results and discussion

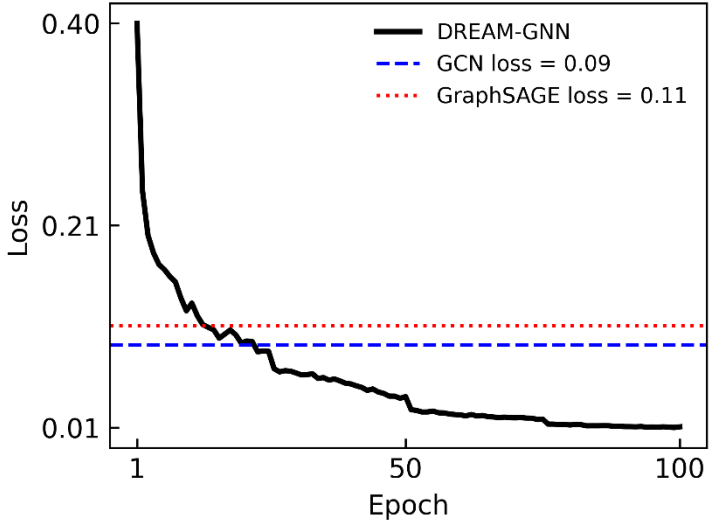

**Fig. 7.** Training-loss trajectory of DREAM-GNN over 100 epochs (solid black). Horizontal reference lines indicate the terminal training losses of the baseline architectures: GCN (0.09, blue dashed) and GraphSAGE (0.11, red dotted). DREAM-GNN converged to a loss of ≈ 0.01, roughly an order of magnitude lower than either baseline – highlighting superior representational capacity compared to established model architectures.

The training-loss trajectory of the DREAM-GNN is shown in Fig. 7, alongside horizontal reference lines that mark the terminal losses of the naïve GCN and GraphSAGE baselines. DREAM-GNN achieved an MSE of ≈ 0.01, while the baselines stabilized at substantially higher values – 0.11 for GraphSAGE and 0.09 for GCN – an order-of-magnitude performance gap in favor of DREAM-GNN. Although the proposed network contained 11.6 million trainable parameters, roughly 15× more than the 0.79 million in each baseline, this additional capacity was well utilized: the edge-aware, multiscale architecture delivered markedly richer representations rather than over-fitting, translating directly into superior predictive accuracy.

Fig. 8 presents a side by side comparison of temperature, pressure, and velocity magnitude fields predicted by the three graph surrogates against the CFD ground truth. The first row, Figs. 8(a)–(c), shows the reference CFD solution: sharp thermal gradients following the contours of the fin surface, a well-defined high-pressure velocity stagnation region on the upstream face, and a coherent recirculation region with accelerated side channel jets downstream. In the second and third rows, the naïve baselines – GCN in Figs. 8(d)–(f) and GraphSAGE in Figs. 8(g)–(i) – captured only the broad trends: temperature contours were blurred, the stagnation pressure pocket was diffuse, the wake velocity was oversmoothed, and boundary layer separation was delayed, with noticeable artifacts in the pressure field. The fourth row, DREAM-GNN in Figs. 8(j)–(l), recovered the fine scale details evident in the CFD: the near wall thermal boundary layer heating in Fig. 8(j), the well-defined, highly-resolved, and precisely captured high to low pressure transition in Fig. 8(k), and the recirculation zone and shear layer jets in Fig. 8(l) aligned closely with their counterparts in Figs. 8(a)–(c). By preserving sharp gradients and eliminating baseline noise, DREAM-GNN demonstrates its capability to reproduce boundary driven, multiscale phenomena – qualitative evidence that supports the order of magnitude error reduction seen in the convergence plot show in Fig. 7.

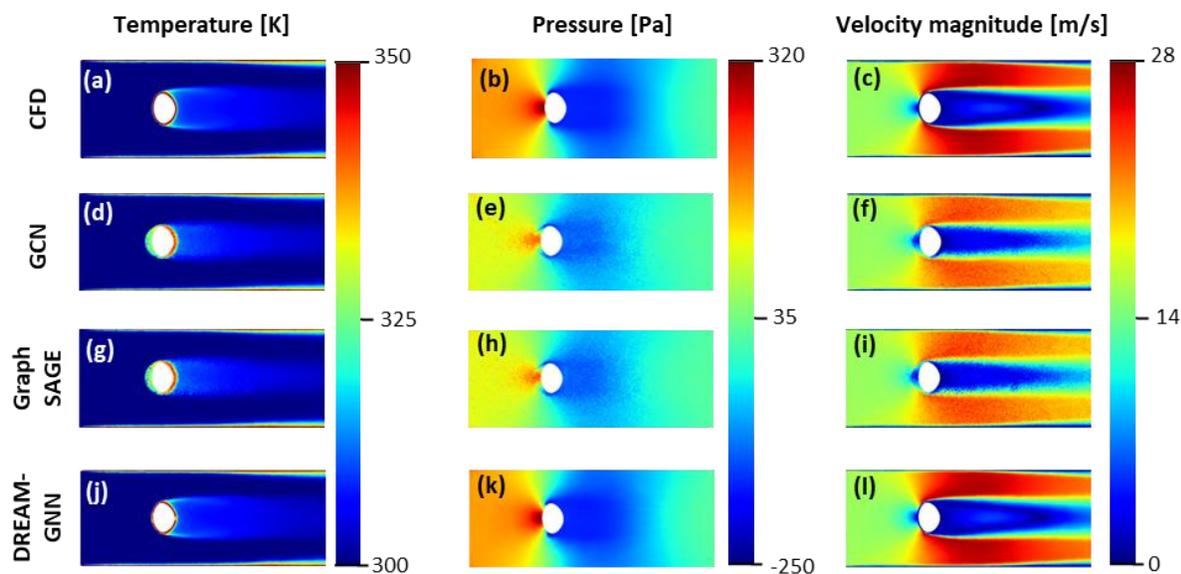

**Fig. 8.** Comparison of predicted temperature, pressure, and velocity magnitude fields for a bluff body fin case. Rows show CFD ground truth (a-c), baseline GNN surrogates (GCN: (d-f), and GraphSAGE: (g-i)), and the proposed DREAM-GNN (j-l). DREAM-GNN closely reproduced boundary layer heating, stagnation pressure, and wake recirculation captured by CFD, while the baselines exhibit blurring and noise in high gradient regions.

Fig. 9 juxtaposes the ground truth CFD solutions with DREAM-GNN predictions for three geometrically distinct arbitrary pin-fin configurations from the test data set. In the first set, panels (a)–(c) exhibit the reference temperature, pressure and velocity fields for the first pin-fin, while panels (d)–(f) display the corresponding DREAM-GNN outputs. The agreement is essentially pointwise: the network preserved the narrow thermal rim that formed along the upstream face, reproduced both the magnitude and spatial extent of the upstream high-pressure stagnation pocket, and captured the length and width of the recirculation zone together with its accelerated side channel jets. This first pin-fin resembled several members of the training ensemble in aspect ratio and blockage ratio, and the model's edge aware encoder retained the steep boundary layer gradients that characterize such fins; multiscale message passing then disseminated this information throughout the short wake, leaving negligible room for perceptible error.

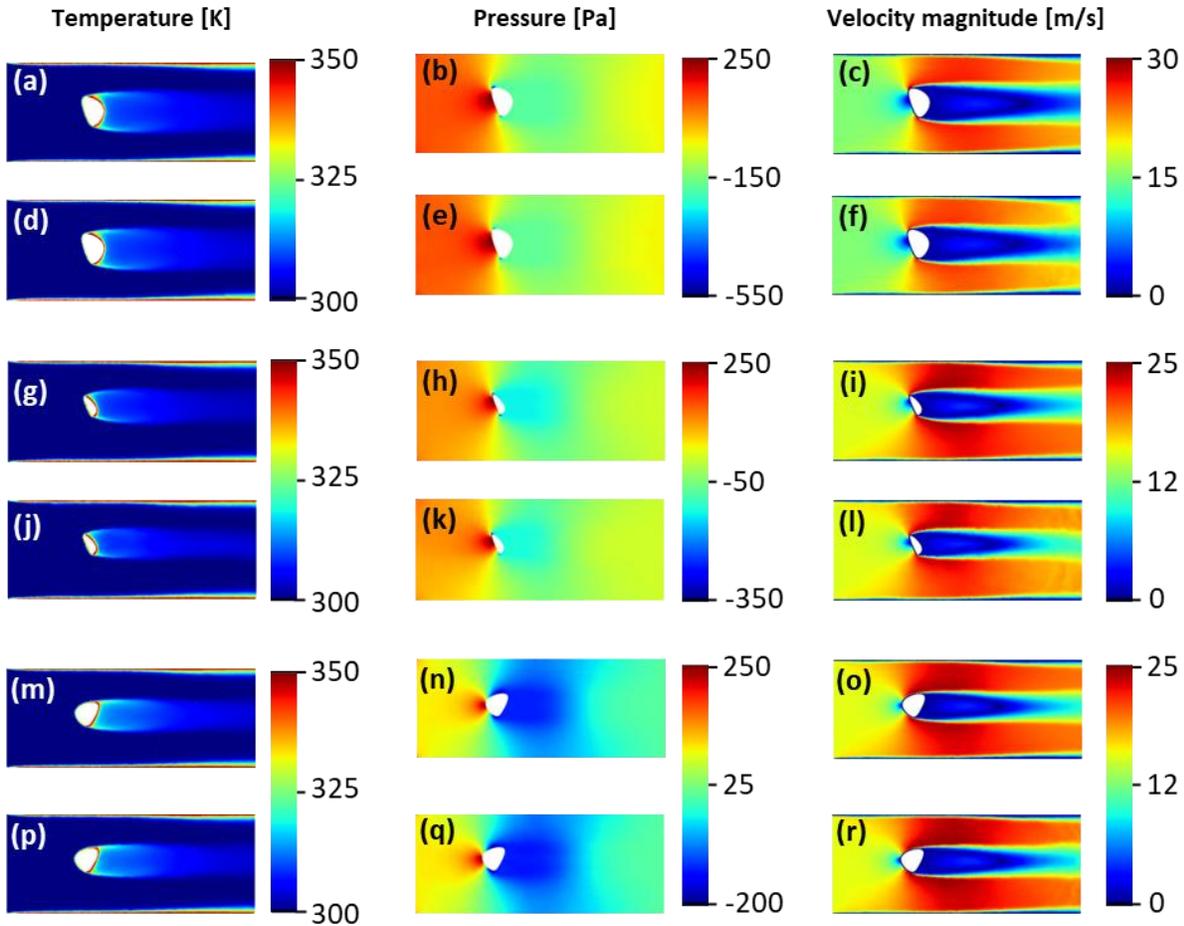

**Fig. 9.** Ground truth CFD fields (top row of each pair) versus DREAM-GNN predictions (bottom row) for three pin-fin geometries. For every fin, the left, center, and right columns display temperature, pressure, and velocity magnitude, respectively. *The first arbitrary pin-fin:* (a–c) CFD and (d–f) DREAM-GNN; *The second arbitrary pin-fin:* (g–i) CFD and (j–l) DREAM-GNN; *The third arbitrary pin-fin:* (m–o) CFD and (p–r) DREAM-GNN. Across all cases DREAM-GNN reproduced the leading-edge heating, stagnation pressure pocket, and wake recirculation seen in the CFD reference, with only minor smoothing – most evident as a slightly shortened low velocity core in panel (l) compared with its ground truth counterpart (i).

The second arbitrary pin-fin, depicted in panels (g)–(i) for the CFD reference and (j)–(l) for DREAM-GNN, posed a sterner test. A sharper nose and higher blockage ratio generated a shorter, narrower wake in the CFD solution. While the network continued to track temperature and upstream pressure with high fidelity, it underpredicted the downstream momentum deficit: the low velocity core in panel (l) is shorter and narrower than its counterpart in panel (i), and the pressure dip in panel (k) recovers slightly too quickly. These discrepancies arise from two interacting factors. First, the loss function weighed all nodes equally; because the recirculation zone occupied a small fraction of the domain, its contribution to the global error budget was limited, encouraging the optimizer to privilege more extensive, lower gradient regions. Second, the effective receptive field of the graph, although enlarged by hierarchical message passing, still contracted with distance from the fin, so far wake nodes received attenuated information about earlier separation and recirculation patterns.

Panels (m)–(o) and (p)–(r) present the third fin, a bluff front design that accentuated leading-edge gradients and narrowed the shear layer. DREAM-GNN accurately reproduced the intensified boundary layer heating, matched peak wall temperatures, and resolved the sharper stagnation pressure transition. The predicted velocity field mirrored the CFD recirculation region and shear layer jets, indicating that the network's skip connections successfully reinjected fine scale information lost during coarsening.

A closer inspection of the four-spline design space clarifies why the "stubby" or "blunt" cross-section embodied by the second arbitrary pin-fin was so sparsely represented in the 1,000-member training set, despite the use of LHS. Each profile was defined by five parameters: a fixed leading-edge radius $r_1 = 1$ mm; three free radii $r_2, r_3$, and $r_4$ drawn independently from the interval $[0.1,1]$ mm; and a global orientation angle $\theta_0$ uniformly distributed over $[0, \pi]$. Once the four polar control points $(r_n, \theta_n + \theta_0)$ with $\theta_n = n\pi/2$ $(n = 0 - 3)$ were set, piecewise cubic splines joined successive points, producing the closed outline shown in Fig. 1.

The streamwise chord $c$ and cross-stream width $d$ of the resulting body are, to first order, the differences between the largest and smallest projections of the radii onto the flow axis and its perpendicular, respectively. Achieving a low chord-to-width ratio $(c/d)$ – the defining geometric trait of the second pin-fin – demands two statistically independent coincidences. First, both radii that ultimately align with the flow direction after rotation must reside in the lower third of their admissible range, compressing $c$. Because LHS placed each $r_k$ once in every decile, the likelihood that a particular pair of radii simultaneously fell into the lowest tertile is $\left(\frac{1}{3}\right)^2 \approx 0.11$. Second, the large, fixed radius $r_1$ must be rotated toward the cross-stream axis so that it inflates $d$. Orientations that satisfy this requirement occupied only a quarter of the $\theta_0$ interval (±22.5° about 90° or 270°), giving a probability of 0.25. The two events are uncorrelated under LHS; therefore, the joint probability of drawing a genuinely stubby fin is the product of the two, $\approx 0.11 \times 0.25 = 0.0275$. On average, then, fewer than thirty of the 1,000 LHS samples inhabit this low-$c/d$ corner of the five-dimensional hyper-cube, whereas hundreds populate the moderate- and high-ratio region typified by the first and the third fins.

This numerical imbalance carried direct consequences for learning. During training, the global mean-squared-error loss was dominated by the plentiful, moderate-slenderness sections; weight updates therefore embedded wake-recovery dynamics characteristic of those shapes. When DREAM-GNN confronted a rare stubby geometry such as the second pin-fin, it reproduced near-body gradients accurately – physics that depends mainly on local curvature, which the network has seen often – but extrapolated the downstream momentum deficit from scant experience, shortening the already brief wake and hastening pressure recovery. In essence, the model behaved as an *approximately Bayesian* estimator whose prior was set by the frequency distribution of the training set: it had a sharp, well-informed prior for moderate $c/d$ fins and a diffuse, weakly informed prior for low-$c/d$ fins, leading to under-prediction of wake length in the latter.

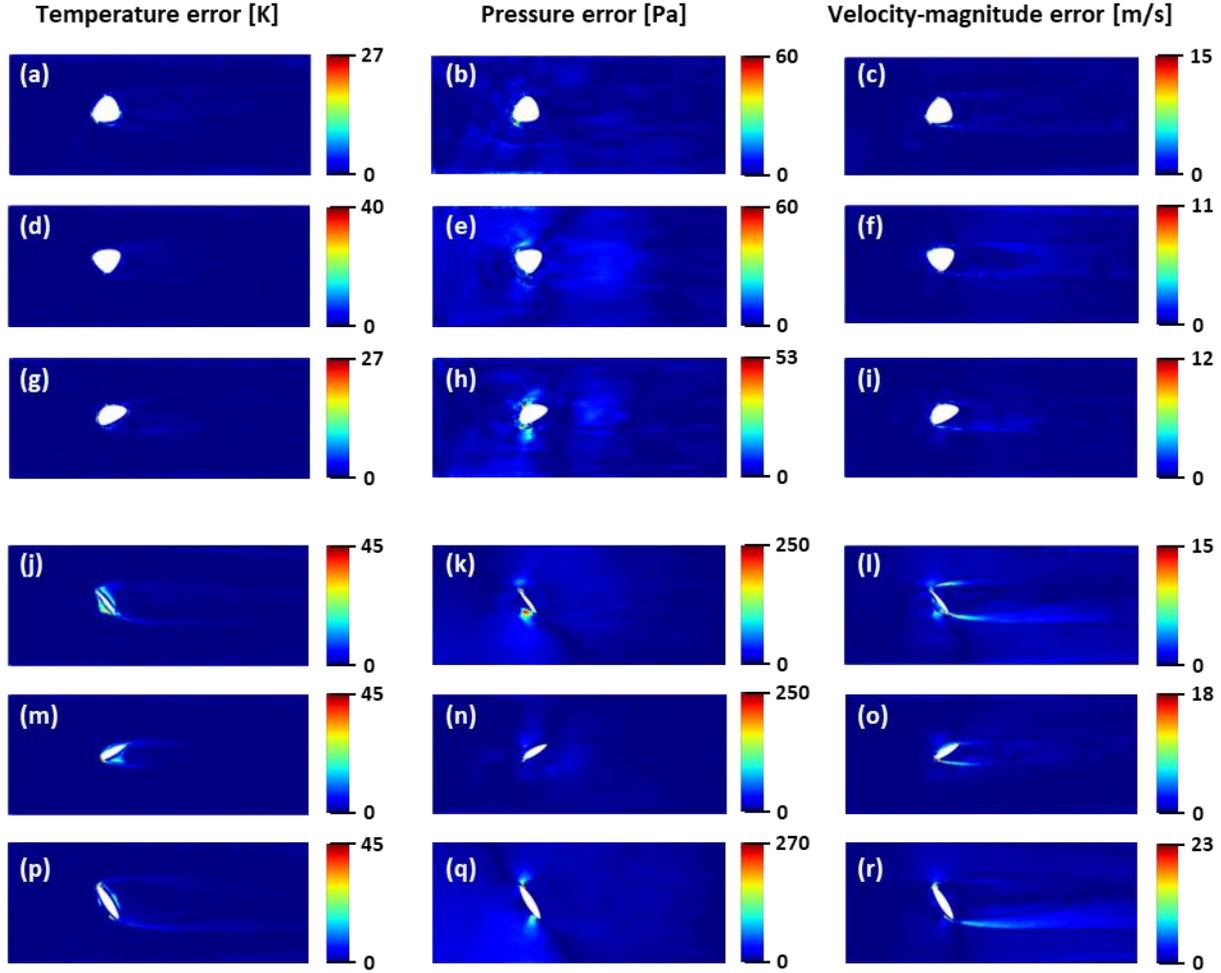

**Fig. 10.** Feature-specific absolute-error maps for three highest-accuracy (rows 1–3) and three lowest-accuracy (rows 4–6) pin-fin geometries. Columns show (left) temperature error in K, (center) pressure error in Pa, and (right) velocity-magnitude error in m/s. Smooth, mid-aspect-ratio fins (rows 1–3) confine pressure error below 60 Pa and velocity-magnitude error below 15 m/s. Fins combining low chord-to-depth ($c/d < 1$) with acute or concave corners (rows 4–6) generate stagnation-point pressure spikes up to 270 Pa; the resulting momentum-deficit mis-prediction propagates downstream as elongated velocity and temperature error plumes.

The variable-specific error maps of Fig. 10 reveal how geometric features drove prediction performance. The first 3 rows correspond to the three best performing fins from the testing dataset; Figs. 10(a)-(c) depict the temperature error, pressure error, and velocity-magnitude error, respectively for the first fin, with the same column ordering used for the rest. Similarly, the last 3 rows indicate the three worst performing fins from the test set. Interestingly, for the three best fins, the stagnation-point pressure error was below 60 Pa and the velocity-magnitude error decayed rapidly as well, keeping temperature errors under 30 K throughout the wake. In contrast, all three worst-performing fins (rows 4–6) shared two features absent from most training samples: a chord-to-depth ratio ($c/d < 1$) and at least one acute or concave leading-edge corner. These traits triggered local pressure overshoots of 250–270 Pa – more than four times larger than in the well-represented shapes – which in turn shortened the recirculation bubble and intensified side-jet velocities. Elevated temperature errors (35–45 K) simply mirrored the mis-placed convective heat paths.

The pattern confirms that DREAM-GNN's largest discrepancies stem from under-represented geometric extremes rather than intrinsic model capacity, suggesting that an active-learning loop focusing on low-$c/d$, high-curvature fins could systematically eliminate the remaining error hotspots.

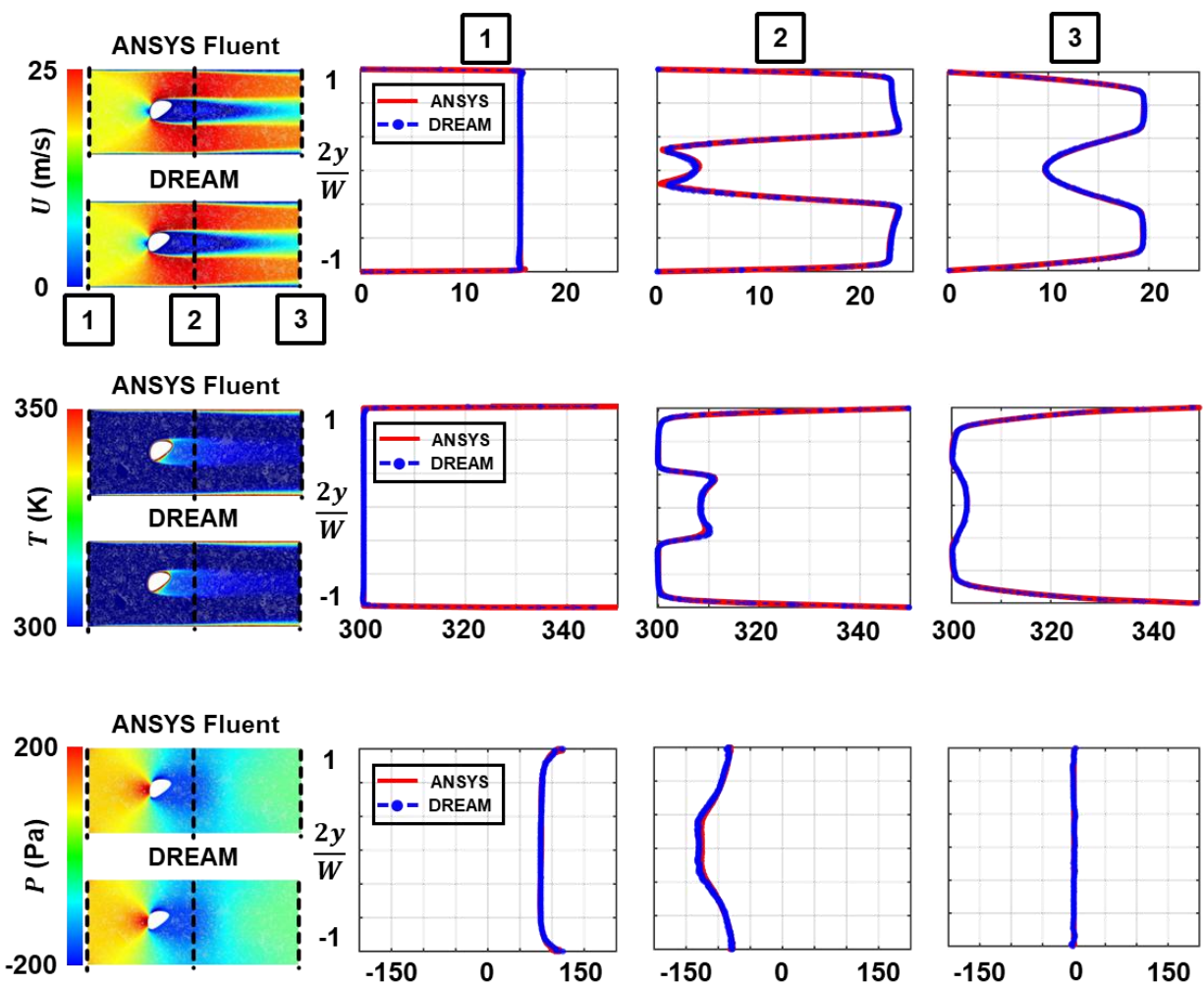

**Fig. 11.** Validation of DREAM predictions against ANSYS Fluent for velocity (top row), temperature (middle row), and pressure (bottom row) profiles in the streamwise direction for an arbitrary pin-fin shape.

To further validate the performance of the DREAM-GNN model in capturing key fluid dynamic behavior, spanwise profiles of velocity magnitude, temperature, and pressure were extracted at three representative streamwise locations: the inlet (1), midline (2), and outlet (3), as shown in Fig. 11. These regions are critical not only for assessing flow development and evaluating thermal-hydraulic performance but also for capturing complex physical phenomena – particularly at the midline (2), where recirculation and boundary layer evolution drive much of the local flow-thermal behavior. Accurate modeling in these regions is essential, as they significantly influence key engineering metrics such as pressure drop and heat transfer, which are often derived from area-weighted averages at critical locations such as the inlet and outlet.

Fig. 11 provides a comprehensive comparison of the velocity magnitude (top row), temperature (middle row), and pressure (bottom row) profiles at these streamwise locations. The respective contour plots from ANSYS Fluent and the DREAM-GNN predictions are displayed in the left-most column for each row. On these contours, a black dotted line is present to indicate the streamwise location (1 – inlet, 2 – midline, 3 – outlet) where these field values were extracted. On the right, the spanwise profiles of the same quantities are shown, with the y-axis normalized using $\frac{2y}{W}$, where $y$ is the vertical coordinate, and $W$ represents the width of the domain. This normalization ensured that the profiles were bound between -1 and 1, enhancing readability. The x-axis represents the magnitude of the respective field variable in the spanwise direction.

The temperature profiles, particularly at the inlet and outlet regions, aligned closely between the DREAM-GNN predictions and ANSYS Fluent, capturing both the thermal boundary layer development and the steady-state behavior at the outlet. The DREAM-GNN also successfully represented the temperature variation near the midline, where recirculation effects are critical in determining thermal performance. Similarly, the velocity profiles showed excellent agreement, with the DREAM-GNN successfully replicating boundary layer growth at the inlet and side walls, boundary layer separation and recirculation in the midline, and the stabilization of the flow at the outlet. The pressure profiles reflect accurate predictions for DREAM-GNN which were consistent with ANSYS Fluent. Notably, at the inlet, where the flow initially contacts the pin-fin, DREAM-GNN captured the stagnation point and the high-pressure region that forms. In the midline, the DREAM-GNN prediction not only replicated the effects of recirculation but also captures the pressure drop associated with the presence of a bluff-body, as depicted by the negative values seen in the midline pressure panel.

In all, the DREAM-GNN model proved to be a robust tool for capturing key fluid dynamic and thermal-hydraulic behavior, outperforming current approaches with unparalleled accuracy across temperature, velocity, and pressure profiles, all while significantly reducing computational time. The DREAM-GNN successfully captured the steep gradients commonly found in complex flow regimes, such as those near the boundary layers, where rapid transitions between high and low values occur. Additionally, the model excelled in handling recirculation zones, where traditional models often struggle to capture the intricate flow behavior, especially in a domain where the shapes under consideration are highly complex and irregular.

## 4. Conclusion

In this study, a Domain-Responsive Edge-Aware Multiscale (DREAM)-GNN was developed for steady, turbulent flow field prediction around complex pin-fins in a two-dimensional channel. The conclusions of this paper can be summarized as follows:

- A training dataset was generated using an automated pipeline which integrated geometry generation, high-resolution meshing, and CFD simulation solutions through ANSYS Fluent. The design space – defined by five geometric parameters – was explored using Latin Hypercube Sampling (LHS) to produce 1,000 different pin-fin configurations. For each design, mesh-based node coordinates, connectivity, and flow solution variables were extracted to construct a rich dataset for training.
- Each CFD simulation was transformed into a graph, in which nodes represented finite-volume cell centers and edges reflected mesh connectivity. Each node carried information about its spatial coordinates, a normalized streamwise position, a one-hot encoded boundary indicator, and a signed distance to the nearest wall, providing geometric and physical context. These graphs served as inputs to the DREAM-GNN, which predicted temperature, velocity magnitude, and pressure at each node based on learned flow physics.
- The DREAM-GNN model accurately captured the steady turbulent flow-thermal behavior across the channel domain. Spanwise profiles of temperature, velocity magnitude, and pressure closely matched results from ANSYS Fluent, successfully reproducing boundary layer development and separation, recirculation effects, and stagnation zones. The model demonstrated strong predictive performance in areas with steep gradients and complex flow patterns, validating its robustness for modeling irregular pin-fin geometries with significantly reduced computational cost.

Ultimately, DREAM-GNN demonstrated a substantial leap in predictive modeling by learning and generalizing the underlying physical behavior of complex, steady turbulent flow regimes. Its ability to accurately capture recirculation, boundary layer development, stagnation, and steep thermal and velocity gradients across highly irregular pin-fin geometries showed that the model internalized core fluid dynamics principles. This data-driven approach, without any explicit physics-based regularizations, effectively captured the same flow-defining features resolved by the RANS solver, achieving performance indistinguishable from ANSYS Fluent, which explicitly solved the underlying partial differential equations of mass, momentum, and energy conservation. As a result, DREAM-GNN offered a powerful and computationally efficient framework for advancing the design and optimization of thermal-fluid systems, especially in cases where traditional GNN models struggled to resolve geometric complexity or required prohibitively expensive simulations.

**Funding**

This material is based upon work supported by the NASA Aeronautics Research Mission Directorate (ARMD) University Leadership Initiative (ULI) under cooperative agreement number 80NSSC21M0068. Any opinions, findings, conclusions, or recommendations expressed in this material are those of the author and do not necessarily reflect the views of the National Aeronautics and Space Administration.

**CRediT authorship contribution statement**

Evan Mihalko and Riddhiman Raut: Writing – original draft, Visualization, Validation, Software, Methodology, Investigation, Formal analysis, Data curation. Amrita Basak: Writing – review & editing, Supervision, Resources, Project administration, Funding acquisition, Conceptualization.

**Data availability statement**

The data presented in this study are available from the corresponding author on reasonable request.

**Declaration of Generative AI and AI-assisted technologies in the writing process**

During this work's preparation, OpenAI ChatGPT has been used to improve grammar and readability.

**Declaration of competing interest**

The authors declare that they have no known competing financial interests or personal relationships that could have appeared to influence the work reported in this paper.